\documentclass[10pt]{article}
\usepackage{amsmath}
\usepackage{amssymb}
\usepackage{graphicx}
\usepackage{fullpage}
\usepackage{float}
\usepackage{algorithm}
\usepackage{algorithmic}
\usepackage{hyperref}
\usepackage{authblk}
\usepackage{breqn}

\author{Vincent Delos$^*$ and Denis Teissandier$^{**}$}
\date{
    University of Bordeaux \\
    CNRS, National Center for French Research \\
    I2M, UMR 5295 \\
    Talence, F-33400, France \\
    $^*$E-mail: \url{v.delos@i2m.u-bordeaux1.fr}\\
    $^{**}$E-mail: \url{d.teissandier@i2m.u-bordeaux1.fr}\\
}

\title{Minkowski sum of polytopes defined by their vertices}

\begin{document}

\maketitle

\begin{abstract}
Minkowski sums are of theoretical interest and have applications in fields related to industrial backgrounds. In this paper we focus on the specific case of summing polytopes as we want to solve the tolerance analysis problem described in \cite{Teissandier-CIRP}. Our approach is based on the use of linear programming and is solvable in polynomial time. The algorithm we developped can be implemented and parallelized in a very easy way.

{\bf keywords:} Computational Geometry, Polytope, Minkowski Sum, Linear Programming, Convex Hull.
\end{abstract}

%
%
\section{Introduction}
Tolerance analysis is the branch of mechanical design dedicated to studying the impact of the manufacturing tolerances on the functional constraints of any mechanical system. Minkowski sums of polytopes are useful to model the cumulative stack-up of the pieces and thus, to check whether the final assembly respects such constraints or not, see \cite{Homri2013} and \cite{Srinivasan1993}.  We are aware of the algorithms presented in \cite{Fukuda20041261}, \cite{Fukuda2005_882}, \cite{Teissandier-hal-00635842} and \cite{Delos-CMCGS} but we believe that neither the list of all edges nor facets are mandatory to perform the operation. So we only rely on the set of vertices to describe both polytope operands. In a first part we deal with a ``natural way'' to solve this problem based on the use of the convex hulls. Then we introduce an algorithm able to take advantage of the properties of the sums of polytopes to speed-up the process. We finally conclude with optimization hints and a geometric interpretation.

\section{Basic properties}

\subsection{Minkowski sums}

Given two sets $A$ and $B$, let $C$ be the Minkowski sum of $A$ and $B$
\begin{center}
  $ C = A + B = \{ c \in \mathbb{R}^n, \exists a \in A, \exists b \in B / c = a+b \} $
\end{center}

\subsection{Polytopes}

A polytope is defined as the convex hull of a finite set of points, called the $\mathcal{V}$-representation, or as the bounded intersection of a finite set of half-spaces, called the $\mathcal{H}$-representation. The Minkowski-Weyl theorem states that both definitions are equivalent.


\section{Sum of $\mathcal{V}$-polytopes}

In this paper we deal with  $\mathcal{V}$-polytopes i.e. defined as the convex hull of a finite number of points. We note $\mathcal{V}_A$, $\mathcal{V}_B$ and $\mathcal{V}_C$ the list of vertices of the polytopes $A$, $B$ and $C=A+B$. We call $\mathcal{V}_C$ the list of \textit{Minkowski vertices}. We note $ k = Card( \mathcal{V}_A) $ and $ l = Card( \mathcal{V}_B) $.

\subsection{Uniqueness of the Minkowski vertices decomposition}

Let $A$ and $B$ be two $\mathbb{R}^n$-polytopes and $\mathcal{V}_A$, $\mathcal{V}_B$ their respective lists of vertices. Let $C = A + B$ and $c = a+b$ where $a \in \mathcal{V}_A$ and $b \in \mathcal{V}_B$.

\begin{equation}
\label{basicprop}
  c \in \mathcal{V}_C \Leftrightarrow \text{the decomposition of $c$ as a sum of elements of $A$ and $B$ is unique}
\end{equation}

We recall that in \cite{Fukuda20041261}, we see that the vertex $c$ of $C$, as a face, can be written as the Minkowski sum of a face from $A$ and a face from $B$. For obvious reasons of dimension, $c$ is necessarily the sum of a vertex of $A$ and a vertex of $B$. Moreover, in the same article, Fukuda shows that its decomposition is unique.

Reciprocally let $ a \in \mathcal{V}_A $ and $ b \in \mathcal{V}_B $ be vertices from polytopes $A$ and $B$ such that $ c = a+b $ is unique. Let $ c_1 \in C $ and $ c_2 \in C $ such as $ c = \frac{1}{2}(c_1 + c_2) = \frac{1}{2}(a_1+b_1 + a_2+b_2) = \frac{1}{2}(a_1 + a_2) + \frac{1}{2}(b_1+b_2) = a+b $ with $ a = \frac{1}{2}(a_1 + a_2) $ and $ b = \frac{1}{2}(b_1+b_2) $ because the decomposition of $c$ in elements from $A$ and $B$ is unique. Given that $a$ and $b$ are two vertices, we have $ a_1 = a_2 $ and $ b_1 = b_2 $ which implies $ c_1 = c_2 $. As a consequence $c$ is a vertex of $C$.

\subsection{Summing two lists of vertices}

Let $A$ and $B$ be two $\mathbb{R}^n$-polytopes and $\mathcal{V}_A$, $\mathcal{V}_B$ their lists of vertices, let $ C = A + B $.

\begin{equation}
  C = Conv( \{ a+b, a \in \mathcal{V}_A, b \in \mathcal{V}_B \} )
\end{equation}

We know that $ \mathcal{V}_C \subset \mathcal{V}_A + \mathcal{V}_B $ because a Minkowski vertex has to be the sum of vertices from $A$ and $B$ so $ C = Conv(\mathcal{V}_C) \subset Conv( \{ a+b, a \in \mathcal{V}_A, b \in \mathcal{V}_B \} ) $.

The reciprocal is obvious as $ Conv( \{ a+b, a \in \mathcal{V}_A, b \in \mathcal{V}_B \} ) \subset Conv( \{ a+b, a \in A, b \in B \} ) = C $ as $ C = A+B $ is a convex set.

\begin{figure}
  \begin{center}
    \includegraphics[scale=0.5]{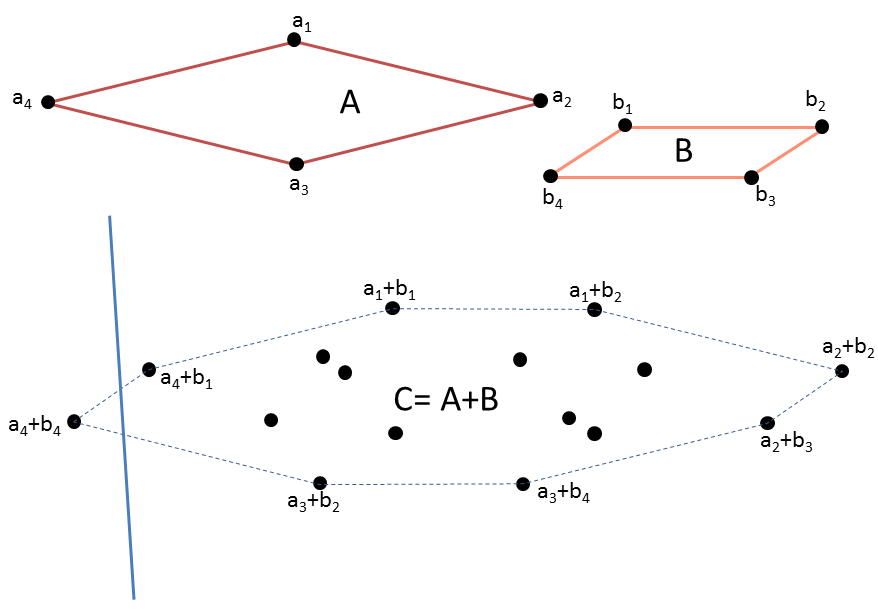}
    \caption{\fontsize{8}{9.6}{Computing the vertices of the sum of two $\mathcal{V}$-polytopes through a convex hull algorithm}}
    \label{vsum}
  \end{center}
\end{figure}

At this step an algorithm removing all points which are not vertices of $C$ from $ \mathcal{V}_A + \mathcal{V}_B $ could be applied to compute $\mathcal{V}_C$. The basic idea is the following: if we can build a hyperplane separating $(a_u+b_v)$ from the other points of $ \mathcal{V}_A + \mathcal{V}_B $ then we have a Minkowski vertex, otherwise $(a_u+b_v)$ is not an extreme point of the polytope $C$. The process trying to split the cloud of points is illustrated in \textbf{Figure} \ref{vsum}.

To perform such a task, a popular technique given in \cite{Fukuda2004_faq} solves the following linear programming system. In the case of summing polytopes, testing whether the point $ (a_u+b_v) $ is a Minkowski vertex or not, means finding $ (\gamma, \gamma_{uv}) \in \mathbb{R}^n \times \mathbb{R} $ from a system of $k \times l$ inequalities:

\begin{equation*}
\left\{ 
\begin{array}{l l}
  < \gamma, a_i+b_j >  - \gamma_{uv} \leq 0 ~; \forall (i,j) \in \{1, ..,k\} \times \{1,..,l\} ~; (i,j) \neq (u,v) \\
  < \gamma, a_u+b_v > - \gamma_{uv} \leq 1 \\
  f^* = \max ( < \gamma, a_u+b_v > - \gamma_{uv} )
\end{array} \right.
\end{equation*}

So if we define the matrix $
 \Gamma =
\begin{pmatrix}
  a_{1,1} + b_{1,1} & \cdots & a_{1,n} + b_{1,n} & -1 \\
  \vdots & \ddots & \vdots & \vdots \\
 a_{k,1} + b_{l,1} & \cdots &  a_{k,n} + b_{l,n} & -1 \\
   a_{u,1} + b_{v,1} & \cdots & a_{u,n} + b_{v,n} & -1
 \end{pmatrix}
 $
 
 then  $
  \Gamma \begin{pmatrix} 
		\gamma \\
		\gamma_{uv}
		\end{pmatrix}
 \leq
	\begin{pmatrix}
	0 \\
	\vdots \\
	0 \\
	1
	\end{pmatrix}
$

The corresponding method is detailed in \textbf{Algorithm} \ref{algbrut}. Now we would like to find a way to reduce the size of the main matrix $\Gamma$ as it is function of the product $ k \times l $.

\begin{algorithm}   
\caption{Compute $C = A+B$ with $A$ and $B$ two $\mathbb{R}^n$-polytopes}
\label{algbrut}     
\begin{algorithmic}
\REQUIRE $A$ $\mathcal{V}$-representation: list of vertices $\mathcal{V}_A$
\REQUIRE $B$ $\mathcal{V}$-representation: list of vertices $\mathcal{V}_B$
\FORALL{$a_u \in \mathcal{V}_A$ and $b_v \in \mathcal{V}_B$}
	\STATE Compute $ f^* = \max ( < \gamma, a_u+b_v > - \gamma_{uv} ) $ with $ \Gamma \begin{pmatrix} 
		\gamma \\
		\gamma_{uv}
		\end{pmatrix} \leq
	\begin{pmatrix}
	0 \\
	... \\
	0 \\
	1
	\end{pmatrix} $, $ \Gamma \in \mathbb{R}^{k \times l} \times \mathbb{R}^{n+1} $
	\IF {$ f^* > 0 $}
		\STATE $ (a_u+b_v) \in \mathcal{V}_C $
	\ELSE
		\STATE $ (a_u+b_v) \notin \mathcal{V}_C $
	\ENDIF
\ENDFOR
\end{algorithmic}
\end{algorithm}

\subsection{Constructing the new algorithm}

In this section we want to use the basic property \ref{basicprop} characterizing a Minkowski vertex. Then the algorithm computes, as done before, all sums of pairs $(a_u,b_v) \in \mathcal{V}_A \times \mathcal{V}_B $ and checks whether there exists a pair $ (a',b') \neq (a_u,b_v) $ with $ a' \in A$, $b' \in B $ such as $ (a'+b') = (a_u+b_v) $. If it is the case then $ (a_u+b_v) \notin \mathcal{V}_C $, otherwise $ (a_u+b_v) \in \mathcal{V}_C $.

$a' = \displaystyle{ \sum_{i=1}^{k} \alpha_i a_i }$ with $ \forall i, \alpha_i \geq 0$ and $\displaystyle{ \sum_{i=1}^{k} \alpha_i } = 1$

$b' = \displaystyle{ \sum_{j=1}^{l} \beta_j b_j }$ with $ \forall j, \beta_j \geq 0$ and $\displaystyle{ \sum_{j=1}^{l} \beta_j } = 1$.

We get the following system:

$ \left\{ 
\begin{array}{l l l l l}
\displaystyle{ \sum_{i=1}^{k} \alpha_i a_i } + \displaystyle{ \sum_{j=1}^{l} \beta_j b_j } = a_u+b_v \\
\displaystyle{ \sum_{i=1}^{k} \alpha_i } = 1 \\
\displaystyle{ \sum_{j=1}^{l} \beta_j } = 1 \\
\forall i, \alpha_i \geq 0 \\
\forall j, \beta_j \geq 0
\end{array} \right.
$

That is to say with matrices and under the hypothesis of positivity for both vectors $\alpha$ and $\beta$:

$
\begin{pmatrix}
  a_{1,1} & a_{2,1} & \cdots & a_{k,1} & b_{1,1} & b_{2,1} & \cdots & b_{l,1} \\
  a_{1,2} & a_{2,2} & \cdots & a_{k,2} & b_{1,2} & b_{2,2} & \cdots & b_{l,2} \\
  \vdots & \vdots  & \ddots & \vdots & \vdots & \vdots & \ddots & \vdots  \\
  a_{1,n} & a_{2,n} & \cdots & a_{k,n} & b_{1,n} & b_{2,n} & \cdots & b_{l,n} \\
  1 & 1 & \cdots & 1 & 0 & 0 & \cdots & 0 \\
  0 & 0 & \cdots & 0 & 1 & 1 & \cdots & 1
 \end{pmatrix}
\begin{pmatrix}
  \alpha_1 \\
  \vdots \\
  \alpha_k \\
  \beta_1 \\
  \vdots \\
  \beta_l
 \end{pmatrix}
 =
\begin{pmatrix}
  a_{u,1} + b_{v,1} \\
  a_{u,2} + b_{v,2} \\
  \vdots \\
  a_{u,n} + b_{v,n} \\
  1 \\
  1
 \end{pmatrix}
$

We are not in the case of the linear feasibility problem as there is at least one obvious solution:

$ p_{u,v} = ( \alpha_1, \cdots, \alpha_k, \beta_1, \cdots, \beta_l ) = ( 0, \cdots, 0, \alpha_u=1, 0, \cdots, 0, 0, \cdots, 0, \beta_v=1, 0, \cdots, 0 ) $

The question is to know whether it is unique or not. This first solution is a vertex $ p_{u,v} $ of a polyhedron in $\mathbb{R}^{k+l}$ that verifies $(n+2)$ equality constraints with positive coefficients. The algorithm tries to build another solution making use of linear programming techniques. We can note that the polyhedron is in fact a polytope because it is bounded. The reason is that, by hypothesis, the set in $\mathbb{R}^{k}$ of convex combinations of the vertices $a_i$ is bounded as it defines the polytope $A$. Same thing for $B$ in $\mathbb{R}^{l}$. So in $\mathbb{R}^{k+l}$ the set of points verifying both constraints simultaneously is bounded too.

So we can write it in a more general form:

$ P \begin{pmatrix}
  \alpha \\
  \beta
 \end{pmatrix} = 
\begin{pmatrix}
  a_u + b_v \\
  1 \\
  1
 \end{pmatrix},
 P \in \mathbb{R}^{n+2} \times \mathbb{R}^{k+l}, \alpha \in \mathbb{R}^{k}_+, \beta \in  \mathbb{R}^{l}_+, a_u \in  \mathbb{R}^{n}, b_v \in  \mathbb{R}^{n}
   $

  where only the second member is function of $u$ and $v$.

It gives the linear programming system:

\begin{equation}
\left\{ 
\begin{array}{l l}
  P \begin{pmatrix}
    \alpha \\
    \beta
  \end{pmatrix} = 
  \begin{pmatrix}
    a_u + b_v \\
    1 \\
    1
  \end{pmatrix} \\
\begin{pmatrix} 
		\alpha \\
		\beta
		\end{pmatrix} \geq 0 \\
  f^* = \max ( 2 - \alpha_u - \beta_v )
\end{array} \right.
\end{equation}

Thanks to this system we have now the basic property the algorithm relies on:

\begin{equation}
  a_u \in \mathcal{V}_A, b_v \in \mathcal{V}_B, (a_u + b_v) \in \mathcal{V}_C \Leftrightarrow f^*=0
\end{equation}

$ f^*=0 \Leftrightarrow $ there exists only one pair $ ( \alpha_u, \beta_v ) = (1, 1) $ to reach the maximum $ f^* $ as $ \sum_{i=1}^{k} \alpha_i = 1 $ and $ \sum_{j=1}^{l} \beta_j = 1 $ $\Leftrightarrow$ the decomposition of $ c = (a_u + b_v) $ is unique $ \Leftrightarrow c \in \mathcal{V}_C $

It is also interesting to note that when the maximum $f^*$ has been reached:

$ \alpha_u = 1 \Leftrightarrow \beta_v = 1 \Leftrightarrow f^*=0 $

\begin{algorithm}   
\caption{Compute $C = A+B$ with $A$ and $B$ two $\mathbb{R}^n$-polytopes}
\label{algopt}     
\begin{algorithmic}
\REQUIRE $A$ $\mathcal{V}$-representation: list of vertices $\mathcal{V}_A$
\REQUIRE $B$ $\mathcal{V}$-representation: list of vertices $\mathcal{V}_B$
\FORALL{$a_i \in \mathcal{V}_A$ and $b_j \in \mathcal{V}_B$}
	\STATE Compute $ f^* = \max ( 2 - \alpha_i - \beta_j ) $ with $ P \begin{pmatrix} 
		\alpha \\
		\beta
		\end{pmatrix} = 
	\begin{pmatrix}
	a_i + b_j \\
	1 \\
	1
	\end{pmatrix} $
        \STATE $ P \in \mathbb{R}^{n+2} \times \mathbb{R}^{k+l} $ and $\begin{pmatrix} 
		\alpha \\
		\beta
		\end{pmatrix} \geq 0 $
	\IF {$ f^* = 0 $}
		\STATE $ (a_i+b_j) \in \mathcal{V}_C $
	\ELSE
		\STATE $ (a_i+b_j) \notin \mathcal{V}_C $
	\ENDIF
\ENDFOR
\end{algorithmic}
\end{algorithm}

\subsection{Optimizing the new algorithm and geometric interpretation}

The current state of the art runs $k \times l$ linear programming algorithms and thus is solvable in polynomial time. We presented the data such that the matrix $P$ is invariant and the parametrization is stored in both the second member and the objective function, so one can take advantage of this structure to save computation time. A straight idea could be using the classical sensitivity analysis techniques to test whether $(a_u + b_v)$ is a Minkowski vertex or not from the previous steps, instead of restarting the computations from scratch at each iteration.

Let's switch now to the geometric interpretation, given $ a \in \mathcal{V}_A $, let's consider the cone generated by all the edges attached to $a$ and pointing towards its neighbour vertices. After translating its apex to the origin $O$, we call this cone $C_O(a)$ and we call $C_O(b)$ the cone created by the same technique with the vertex $b$ in the polytope $B$.

The method tries to build a pair, if it exists, $ (a',b') $ with $ a' \in A$, $b' \in B $ such that $ (a+b) = (a'+b') $. Let's introduce the variable $ \delta = a'-a =b-b' $, and the straight line $\Delta = \{ x \in \mathbb{R}^n : x = t \delta, t \in \mathbb{R} \} $.

So the question about $ (a+b) $ being or not a Minkowski vertex can be presented this way:
\begin{equation}
  a \in \mathcal{V}_A, b \in \mathcal{V}_B, (a+b) \notin \mathcal{V}_C \Leftrightarrow \exists \Delta = \{ x \in \mathbb{R}^n : x = t \delta, t \in \mathbb{R} \} \subset C_O(a) \cup C_O(b)
\end{equation}

\begin{figure}
  \begin{center}
    \includegraphics[scale=0.5]{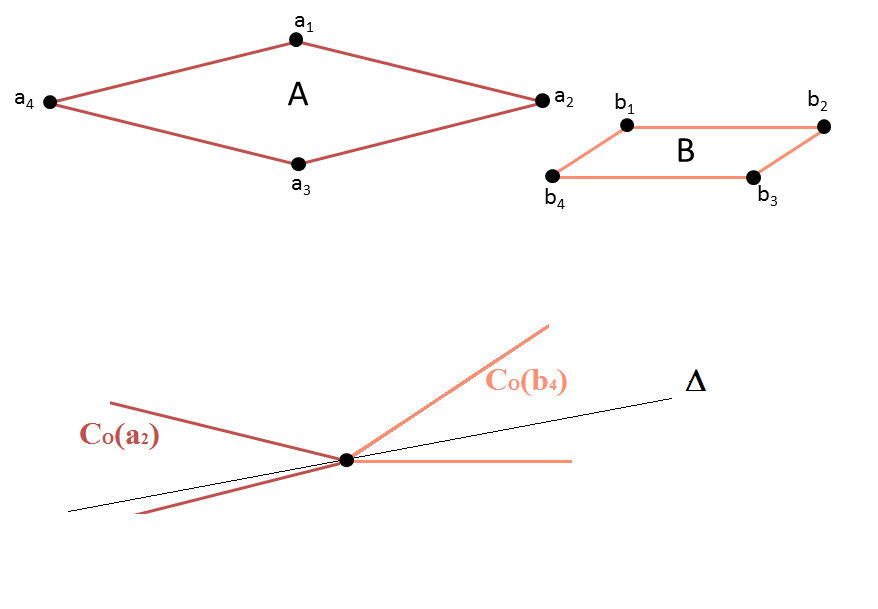}
    \caption{\fontsize{8}{9.6}{$(a_2 + b_4)$ is not a vertex of C=A+B as $\Delta \subset C_O(a_2) \cup C_O(b_4)$}}
    \label{vsum2}
  \end{center}
\end{figure}

The existence of a straight line inside the reunion of the cones is equivalent to the existence of a pair $(a',b')$ such that $ (a+b) = (a'+b') $ which is equivalent to the fact that $ (a'+b') $ is not a Minkowski vertex. This is illustrated in \textbf{Figure} \ref{vsum2}. The property becomes obvious when we understand that if $ (a',b') $ exists in $ A \times B $ then $ (a'-a) $ and $ (b'-b) $ are symmetric with respect to the origin. Once a straight line has been found inside the reunion of two cones, we can test this inclusion with the same straight line for another pair of cones, here is the geometric interpretation of an improved version of the algorithm making use of what has been computed in the previous steps.

We can resume the property writing it as an intersection introducing the cone $-C_O(b)$ being the symmetric of $C_O(b)$ with respect to the origin.

\begin{equation}
\label{primcone}
   a \in \mathcal{V}_A, b \in \mathcal{V}_B, (a+b) \in \mathcal{V}_C \Leftrightarrow C_O(a) \cap -C_O(b) = \{O\}
\end{equation}

\section{Conclusion}

In this paper, our algorithm goes beyond the scope of simply finding the vertices of a cloud of points. That's why we have characterized the Minkowski vertices. However, among all the properties, some of them are not easily exploitable in an algorithm. In all the cases we have worked directly in the polytopes $A$ and $B$, i.e. in the primal spaces and only with the polytopes $\mathcal{V}$-descriptions. Other approaches use dual objects such as normal fans and dual cones. References can be found in \cite{Teissandier-hal-00635842},  \cite{Delos-CMCGS} and \cite{Weibel3883} but they need more than the $\mathcal{V}$-description for the polytopes they handle. This can be problematic as obtaining the double description can turn out to be impossible in high dimensions, see \cite{Fukuda20041261} where Fukuda uses both vertices and edges. Reference \cite{Teissandier-hal-00635842} works in $ \mathbb{R}^3 $ in a dual space where it intersects dual cones attached to the vertices, and it can be considered as the dual version of property \ref{primcone} where the intersection is computed with primal cones. It actually implements Weibel's approach described in \cite{Weibel3883}. Such a method has been recently extended to any dimension for $\mathcal{H}\mathcal{V}$-polytopes in \cite{Delos-CMCGS}.

\section{Special thanks}

We would like to thank Pr Pierre Calka from the LMRS in Rouen University for his precious help in writing this article.

\end{document}